\documentclass[a4paper,11pt]{article}
\usepackage[utf8]{inputenc}
\usepackage{graphicx}
\usepackage{xfrac}
\usepackage{amsmath,amsfonts,amssymb}
\usepackage{subcaption}
\usepackage{color}
\usepackage{slashed}
\usepackage{cite}
\usepackage{csquotes}
\usepackage{authblk}
\usepackage[normalem]{ulem}
\usepackage[colorlinks=true, urlcolor=blue, linkcolor=blue, citecolor=blue]{hyperref}
\usepackage[toc,page]{appendix}


\title{\bf Graviballs and Dark Matter}
\author[1]{B. Guiot\thanks{benjamin.guiot@usm.cl}}
\author[1]{A. Borquez\thanks{andres.borquez.14@usmcl.onmicrosoft.com}}
\author[2]{A. Deur\thanks{deurpam@jlab.org}}
\author[3]{K. Werner\thanks{werner@subatech.in2p3.fr}}
\affil[1]{Departamento de F\'isica, Universidad T\'ecnica Federico Santa Mar\'ia; Casilla 110-V, Valparaiso, Chile}
\affil[2]{University of Virginia, Charlottesville, Virginia 22901, USA}
\affil[3]{SUBATECH, Universit\'e de Nantes, IMT Atlantique, IN2P3/CNRS, 4 rue Alfred Kastler, 44307 Nantes cedex 3, France}

\date{}

\begin{document}

\maketitle

\begin{abstract}
We investigate the possible existence of graviballs, a system of bound gravitons, and show that two gravitons can be bound together by their gravitational interaction. This idea connects to black hole formation by a high-energy $2\to N$ scattering and to the gravitational geon studied by Brill and Hartle. Our calculations rely on the formalism and techniques of quantum field theory, specifically on low-energy quantum gravity. By solving numerically the relativistic equations of motion, we have access to the space-time dynamics of the (2-gravitons) graviball formation. We argue that the graviball is a viable dark matter candidate and we compute the associated gravitational lensing. 
\end{abstract}

\newpage

\tableofcontents

\section{Introduction}
While the origin of the missing mass in galaxy and in galaxy cluster
continue to be actively studied, the existence of large amount dark matter (DM)
made of yet-to-be-identified particles has emerged as the leading explanation. Under this hypothesis, the near-consensus is that these particles are non-relativistic (cold DM). The main argument against hot DM particles -- that their small mass, i.e. relativistic nature, prevents them to make the small initial overdensities necessary for structures to form fast enough -- applies even more to massless particles. However, that argument is not systematically applicable, as exemplified by the case of axions as viable cold DM candiates despite of their diminutive mass. Other examples of well-localized bound systems of massless or nearly massless components exist: in the context of Quantum Chromodynamics (QCD), an especially instructive case is that of glueballs formed by massless gluons. Another well-known example is the nucleon, in which gluons binding
quarks of negligible mass are also confined despite being strictly massless. In fact, in the massless quark limit, the nucleon is still stable with nearly the same mass, 0.88 GeV rather than 0.94 GeV \cite{PrMuWo}. In this article, we discuss another case where the hot dark matter argument does not apply, and study the possible relation between the graviton and the missing mass problem.\\

It is interesting to draw a parallel between nucleons and galaxies. The mass of these objects is much larger than the sum of their (known) component masses. In the case of a nucleon, it is known that 99\% of its mass is binding energy from the strong interaction, a result in which gluon self-interaction plays a central role. Similarly to gluons, gravitons self-interact. In Refs.~\cite{Deur}, lattice calculations were used to find the influence of gravity's self-interaction on galaxy dynamics, and found that it alleviates the missing-mass problem. However, these calculations were based on the Einstein-Hilbert Lagrangian and obtained in the classical limit. Hence, the results are within the classical
framework of General Relativity (GR), without explicit link to gravitons (except as a vocabulary convenience for the QCD-GR parallel).\\

Our main goal is to investigate the possible existence, and the space-time dynamics, of an object made of gravitons bound together by their gravitational interactions. By analogy with QCD's glueballs, we call these objects graviballs. The idea relates to the gravitational geon\footnote{Geons are discussed in appendix \ref{apgeon}} \cite{BriHar,AndBri}, the classical equivalent of the graviball. This idea has also been studied within quantum field theory and string theory, see \cite{DvaGom,DvGoIs,AdBiVe} and references therein. In these studies, it is shown that the scattering of two gravitons with ultra-Planckian energies is dominated by the production of a large number of small-energy gravitons, resulting in the formation of a black-hole\footnote{In these studies, the black-hole fermionic number is zero. We prefer to use the term graviball to avoid confusion with usual black-hole resulting from the collapse of a massive star.}. The similarities and differences between Refs.~\cite{DvaGom,DvGoIs,AdBiVe} and the present work is discussed in more details in Sec.~\ref{secex}. If graviballs are indeed possible, they would be an evident solution to the missing mass problem. This candidate is more natural than beyond-the-standard-model particles in the sense that no new physics is postulated. Indeed, all the necessary tools for our study have already been developed for other purposes, and our calculations require no additional hypothesis nor free parameters. These tools, those of effective field theory of gravity (or low-energy quantum gravity)\cite{Dono1,Dono2,Burg} and semi-classical calculations, have proven to give accurate results. Here, the term semi-classical refers to calculations based on the potential energy, extracted from scattering amplitudes, see Secs.~\ref{secpot} and \ref{secgrav}.\\

In this article, we start with the simulation of the $2\rightarrow 2$ graviton scattering, presented in Sec.~\ref{secres}. We use semi-classical calculations similar  to  those  performed  in  Ref.~ \cite{BjDoHo} for  the  bending  of  light  by  the  sun, and  solved  numerically  the  relativistic  equations  of  motion (\ref{releq1}-\ref{releq4}). The interest of the simulation is the access to the space-time dynamics of the scattering. We will discuss the cases of $2\rightarrow N$ and $N'\rightarrow N$ scatterings in Sec. \ref{secreal}, where $N\geq N'\gg 1$ are the numbers of gravitons involved in the scattering. However, the implementation of these higher order effects in our dynamical simulation is a technical but non-trivial problem that will be addressed in a future publication. Based on \cite{DvaGom,DvGoIs,AdBiVe} and on general arguments, we will argue in Sec.~\ref{secdis} that the general conclusion obtained for the $2\rightarrow 2$ scattering will not change in situations involving a large number of gravitons. Then, the goals of this article are: 1) to introduce the idea of a possible DM candidate made of gravitons; 2) to present the formalism and the dynamical treatment of the $2\rightarrow 2$ graviton scattering and 3) to discuss the physics and some details of a more realistic simulation involving a large number of gravitons. For the case considered here, we will find that two gravitons can form a bound system of small size. It is indeed generally accepted that the collision of two sufficiently energetic particles will form a black hole. The reason for this is simply that in general relativity, a black hole is formed if a source of energy is located within its Schwarzschild radius. In Sec.~\ref{secdis}, the graviball as a potential solution to the missing mass problem is discussed.\\

 Throughout this article, we use the signature $(+,-,-,-)$ for the flat metric $\eta_{\mu\nu}$, and $\hslash=c=1$.

\section{Black hole formation by ultra-Planckian $2\rightarrow N$ scattering}\label{secex}
In this section, we discuss the results of Refs.~\cite{DvGoIs,AdBiVe} that are closely related to our study. We will underline the similarities and differences with our work, and show their complementarity. The question addressed in Refs.~\cite{DvGoIs,AdBiVe} --namely to find a microscopic description of a black hole agreeing with the general relativity expectation that two colliding ultra-Planckian particles must create such object-- connects to our work on graviballs.\\

Clearly, macroscopic objects such as black holes contain a large number, $N$, of soft gravitons. We are then interested by the scattering $2\rightarrow N$, with $N\gg 1$. Ref.~\cite{DvGoIs} showed that the tree-level amplitude for this process is dominant in the classicalization limit, where the two incoming particles have a center of mass energy $\sqrt{s}$, and the $N$ outgoing particles have momenta proportional to $\sqrt{s}/N$. This limit is also known as the multi-Regge limit or kinematics \cite{Lip}. The study of high-energy scatterings is interdisciplinary, and black hole formation shows some similarities with the study of quark-gluon plasma or that of the nucleon probed at small Bjorken x\cite{BeHeMa}. In \cite{DvGoIs}, the authors found that the perturbative $2\rightarrow N$ amplitude is suppressed by a factor $e^{-N}$, apparently contradicting the classical expectation that a black hole (made of $N$ particles) should be formed with a probability of one. The authors proposed that the perturbative amplitude should be supplemented by a non-perturbative factor $e^N$, corresponding to the black hole entropy.\\

\begin{figure}[!h]
\center
\includegraphics[width=7.0cm]{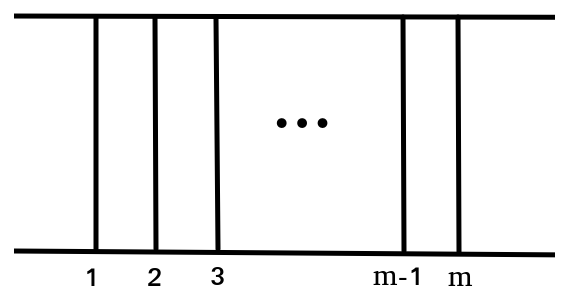}
\caption{S-channel ladder with the exchange of $m$ t-channel gravi-reggeons (gravitons).\label{grib}}
\end{figure}

\begin{figure}[!h]
\center
\includegraphics[width=7.0cm]{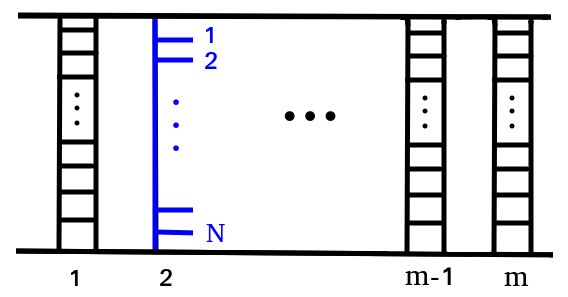}
\caption{Same as Fig. \ref{grib} with reggeized gravitons (or gravi-reggeon). In blue, a cut gravi-reggeon contributing to the inelastic production of particles.\label{cutgrav}}
\end{figure}

Ref. \cite{AdBiVe} is of special interest for this article, because the formalism, detailed in \cite{AmCiVe1,Ven1,Ven2}, can be accommodated to our semi-classical calculations. Indeed, the main idea is that at sufficiently large impact parameter, $b\gg R_s$ with $R_s$ the Schwarzschild radius, the scattering of two ultra-Planckian gravitons is elastic and happens through the exchange of a large number of soft and nearly on-shell gravitons, see Fig. \ref{grib}. This is the microscopic picture  underlying the interaction of two particles via a long-range potential, i.e.  underlying the semi-classical approach. This large number of interactions is similar to what happens in the Gribov-Regge theory \cite{Gribov}. The two energetic gravitons exchange a large number of t-channel objects called gravi-reggeons, see Fig. \ref{grib}.
At large impact parameter $b$, a gravi-reggeon corresponds to the graviton exchanged in the tree-level 4-graviton amplitude Eq.~(\ref{Atree}), and the resummation of these gravi-reggeons is necessary to avoid unitarity violation. At smaller impact parameter the graviton reggeizes and can be cut, resulting in an inelastic production of $N$ particles, see Fig. \ref{cutgrav}.
Note that in the context of heavy-ion collisions, the Gribov-Regge theory has been implemented in several event generators, e.g. EPOS \cite{epos}. In \cite{AdBiVe}, the tree-level $2\rightarrow N$ amplitude of Ref. \cite{DvGoIs} is studied, taking into account the possibility for any number of soft virtual corrections, with energy $E_v<\Lambda$, and any number of soft emissions, with individual energy $E_s<\bar{E}$, where $\Lambda$ and $\bar{E}$ are the cutoffs for the virtual and soft emissions, respectively. For $\bar{E}\gg T_H=\hslash/R_s$, where $T_H$ is the Hawking temperature, the authors found that the amplitude $2\rightarrow N$ is suppressed by a factor of $e^{-N}$, in agreement with \cite{DvGoIs}. In the opposite regime $\bar{E}\ll T_H$, the transition probability can be of order one\footnote{This is true if $\Lambda\sim T_H$, otherwise the amplitude is also exponentially suppressed. With this choice, the difference between the energy cut-offs for virtual and real emissions is large $\bar{E}\ll \Lambda \sim T_H$ and $\bar{E}$ goes dangerously close to zero, potentially explaining the enhancement of the amplitude.}, without requiring the non-perturbative contribution of Ref.~\cite{DvGoIs}, $e^N$, corresponding to the black hole entropy.\\

A first difference between Refs.~\cite{DvGoIs,AdBiVe} and our study is that our goal is more restricted  since it is sufficient to know whether the amount of energy, corresponding to the two incoming gravitons, will be trapped within a sphere of radius $R_s$ or not. For our purpose, it is unnecessary to understand the detailed features of the black hole/graviball such as its spin or entropy. Another difference is that we do not consider only ultra-Planckian, but also Planckian energies. In this case, the graviball is not a macroscopic object. Finally, the main difference and one of the interests of our work is the access to the space-time dynamics, by solving numerically the relativistic equations of motion. The large number of gravitons produced in the 2-graviton scattering are not necessarily produced at the same time; some of them could for instance originate from a gravi-strahlung process, studied in \cite{GruVen,CiCoVe,CiCoCo}. There is then an interesting space-time dynamics which can be explored with the help of numerical simulations. However, the full simulation including the production of a large amount of particles requires the use of advanced Monte-Carlo techniques and will be addressed in a future work. The simulation of this more complex case is discussed in Sec. \ref{secreal}.

\section{Low-energy quantum gravity and semi-classical calculations}\label{secpot}
Non-renormalizable effective theories, e.g. chiral effective field theory \cite{MaEn}, are commonly used. Despite their name, they are renormalizable in a more general sense and can make accurate predictions. The effective Lagrangian is organized in an energy expansion
\begin{equation}
\mathcal{L}_{\text{eff}}=\mathcal{L}_0+\mathcal{L}_1+\mathcal{L}_2+...,
\end{equation}
where $\mathcal{L}_j$ is suppressed by powers of the small energy ratio $E/\Lambda_{\text{HE}}$ compared to $\mathcal{L}_{i}$, $i<j$. Here, $E$ is the low energy scale characterizing the problem while $\Lambda_{\text{HE}}$ is a high energy scale. At this scale and beyond, the effective theory becomes inapplicable. It is usual to consider that the Standard Model of particle physics corresponds only to the $\mathcal{L}_0$ of the true high energy theory, and physics beyond the Standard Model can be  studied with the higher-order Lagrangians $\mathcal{L}_{i>0}$. 

\begin{figure}[!h]
\center
\includegraphics[width=6.0cm]{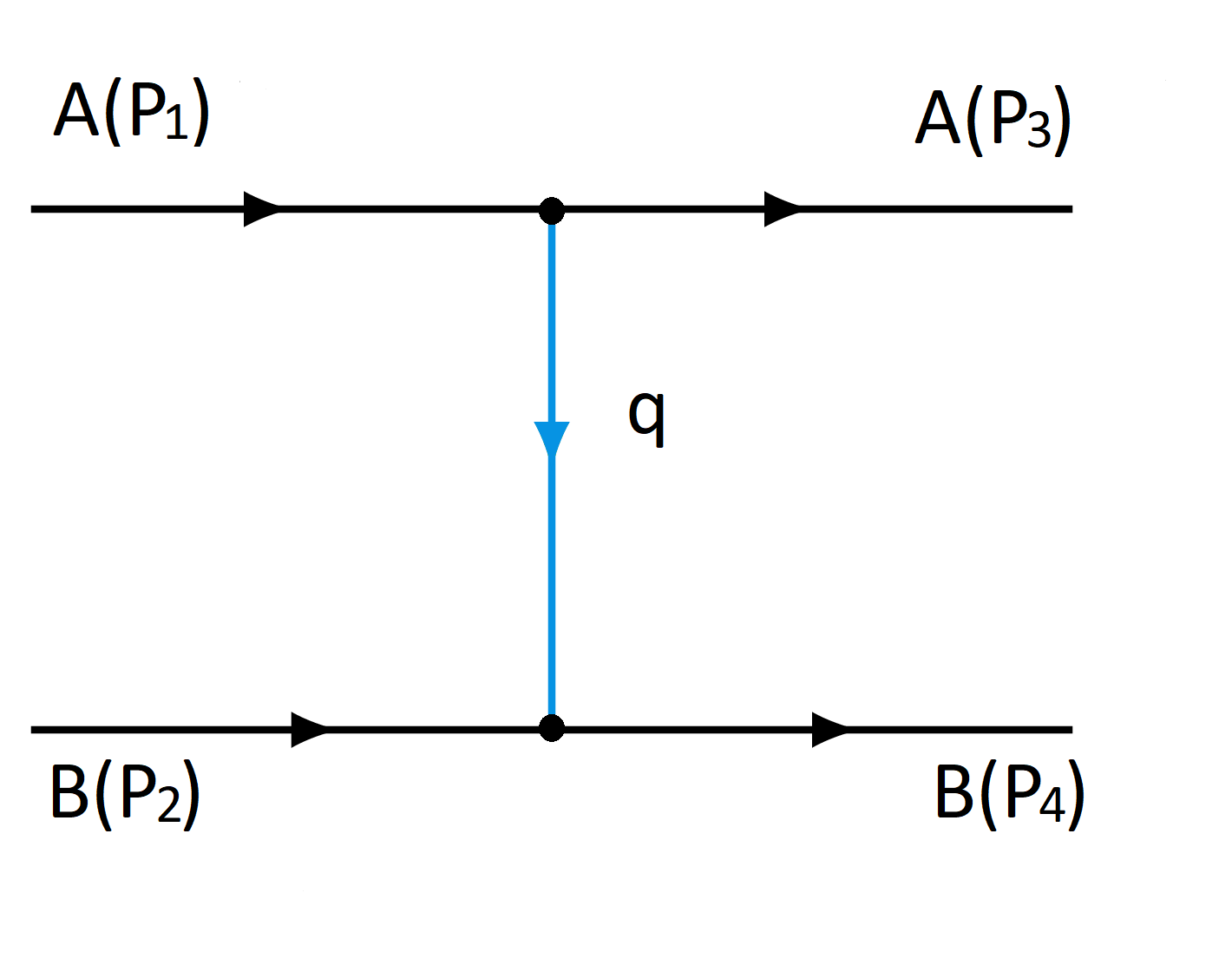}
\caption{
Tree-level diagram for the scattering of particles $A$ and $B$ in the $t$-channel. The $P_i$ denote the particle 4-momenta and the Mandelstam variable $t$ is defined by $t=q^2$.\label{cin}}
\end{figure}

Each $\mathcal{L}_i$ must respect the symmetries of the theory: in the case of gravitation, they must be invariant under general coordinate transformations. For pure gravity\footnote{Pure gravity includes only the gravitational field, i.e. no other bosons, nor fermions.}, the most general Lagrangian reads \cite{Dono1}
\begin{equation}
\frac{\mathcal{L}_{\text{eff}}}{\sqrt{-g}}=\Lambda+\frac{2}{\kappa^2}R+c_1R^2+c_2R_{\mu\nu}R^{\mu\nu}+...,\label{leff}
\end{equation}
where $\kappa^2=32\pi G$, $G$ is the Newton constant, $\Lambda$ is the cosmological constant, $g=\text{det}g_{\mu\nu}$ and $g_{\mu\nu}$ is the metric. The Riemann curvature tensor $R_{\mu\nu\alpha\beta}$, the Ricci tensor $R_{\mu\nu}$ and the Ricci scalar $R$ involve two derivatives of the metric. In momentum space, these derivatives are associated to the 4-momentum, $q$, transferred during the interaction of one graviton with another, see Fig.~\ref{cin}, and Eq. (\ref{leff}) is an expansion in powers of $q$. In this equation the dots denote higher powers of $R$, $R_{\mu\nu}$ and $R_{\mu\nu\alpha\beta}$.  The cosmological constant is generally neglected due to its small experimental value. The Einstein-Hilbert action,
\begin{equation}
\mathcal{S}_{\text{EH}}=\int d^4x \sqrt{-g}\frac{2}{\kappa^2}R,
\end{equation}
corresponds to the lowest order term in Eq. (\ref{leff}) (neglecting $\Lambda$), and provides accurate results for small values of the curvature, $\kappa^2c_1 \, R^2 \ll R$. Eq. (\ref{leff}) shows how general relativity can be treated as any effective field theory. The constants $c_i$ are not predicted by the theory and must be extracted by measurements. At a given order in the energy expansion, there is always a finite number of these constants, and the effective field theory can still make predictions if the number of observables is larger than the number of constants to be determined. 
The gravitational field $h^{\mu \nu}$ is defined as a perturbation  about the background metric $\bar{g}^{\mu \nu}$
\begin{equation}
g^{\mu \nu}=\bar{g}^{\mu \nu}+\kappa h^{\mu \nu}.
\end{equation}
Its quantization involves in particular a gauge fixing term 
and ghosts, see Ref. \cite{Dono1}.\\

The Ultra-Violet (UV) divergences due to loop calculations performed with the Einstein-Hilbert action are of higher order in the energy expansion, and can consequently be absorbed in the renormalization of the constants $c_i$, see Ref. \cite{Dono3} and references therein. As an example, consider the UV divergent part of the one-loop effective Lagrangian, including matter loops~\cite{Dono3}:
\begin{multline}
\Delta \mathcal{L}_{\text{div}}=\frac{1}{16\pi^2}\left(\frac{1}{\epsilon_{\text{uv}}}+\ln 4\pi-\gamma \right)\\
\times \left[\left(\frac{1}{120}R^2+\frac{7}{20}R_{\mu\nu}R^{\mu\nu}\right)+\frac{1}{240}(2R_{\mu\nu}R^{\mu\nu}+R^2) \right],\label{lagdiv}
\end{multline}
where $\gamma$ is the Euler constant. As expected, this Lagrangian is of higher order in the curvature compared to the Einstein-Hilbert Lagrangian. An inspection of Eqs. (\ref{leff}) and (\ref{lagdiv}) shows that the UV divergence can be absorbed in the renormalization of the $c_1$ and $c_2$ constants:
\begin{align}
    c_1^{\overline{MS}}&=c_1+\frac{1}{16\pi^2}\left(\frac{1}{\epsilon_{\text{uv}}}+\ln 4\pi-\gamma \right)\left(\frac{1}{120}+\frac{1}{240}\right),\\
    c_2^{\overline{MS}}&=c_2+\frac{1}{16\pi^2}\left(\frac{1}{\epsilon_{\text{uv}}}+\ln 4\pi-\gamma \right)\left(\frac{1}{120}+\frac{7}{20}\right)
\end{align}
where $\overline{MS}$ indicates that the modified minimal subtraction scheme is used. The loop corrections also contain non-analytic terms of the form $\ln(-q^2)$ or $(q^2)^{-1}$ which cannot be absorbed in the higher-order terms proportional to $q^n$ with $n>0$. Consequently, they are separate from the $c_i$ and genuine predictions of the low-energy effective theory. One of the advantages of effective field theories is this clear separation of high and low energies. In our case, we have a separation of large and small momentum transfers, corresponding to small and long distances. When studying graviballs, it is useful to remember 
that at the event horizon of a black hole, the curvature is still sufficiently small for the application of semi-classical calculations. Then, the dynamics of a particle captured by a black hole can be described to good approximation by the Einstein-Hilbert action, neglecting higher orders in Eq. (\ref{leff}). However, when the particle comes too close to the black hole center, the unknown high energy theory is required. The similar limit of our calculations will be discussed 
at the end of Sec. \ref{secres}.\\

Semi-classical calculations are based on the long-distance potential involved in the scattering of two particles. It is given by the Fourier transform of the non-analytic part of the amplitude $\mathcal{M}(\vec{q})$\cite{HoRo,Kie}
\begin{equation}
V(r)=-\int \frac{d^3\vec{q}}{(2\pi)^3} e^{i\vec{r}.\vec{q}}\frac{\mathcal{M}(\vec{q})}{\sqrt{2E_1}\sqrt{2E_2}\sqrt{2E_3}\sqrt{2E_4}}, \label{pot}
\end{equation}
with $E_i$ the energy of particle $i$ participating to the process, see Fig. \ref{cin}.
In Eq. (\ref{pot}), the long-distance limit $t=q^2\approx -\vec{q}\,^2\rightarrow 0$ is used. 
The distance between the two particles is denoted $r=|\vec{r}|$.\\

As a first example, we consider the muon--electron interaction $\mu^-e^- \rightarrow \mu^-e^-$.
At leading order, there is only one Feynman diagram with a photon exchanged in the $t$ channel. The amplitude is
\begin{equation}
\mathcal{M}^{\text{tree}}=e^2 \bar{u}(P_3,s_3)\gamma^{\mu}u(P_1,s_1)\frac{\eta_{\mu\nu}}{q^2}\bar{u}(P_4,s_4)\gamma^{\nu}u(P_2,s_2),
\end{equation} 
where $e$ is the 
elementary electric charge and the 4-vector $s_i$ describes the spin of particle $i$. The 4-momenta are shown in Fig. \ref{cin} and in this example, the electron corresponds to particle $A$. In the non-relativistic limit, $\mathcal{M}$ is dominated by 
\cite{peskin} 
\begin{equation}
\bar{u}(P',s')\gamma^{0}u(P,s)\approx 2m\delta_{s's}, 
\end{equation}
where $m$ is the fermion mass. Since in this limit $\sqrt{2E_1}\sqrt{2E_3}=2m_e$ and $\sqrt{2E_4}\sqrt{2E_2}=2m_{\mu}$, with $m_e$ and $m_\mu$ the electron and muon masses, respectively, we obtain the repulsive Coulomb potential,
\begin{equation}
V(r)^{\text{tree}}=\int \frac{d^3\vec{q}}{(2\pi)^3} e^{i\vec{r}.\vec{q}}\frac{e^2}{\vec{q}\,^2}=\frac{e^2}{4\pi r}.
\end{equation}
\\

Another example is given in Ref. \cite{BjDoHo}, where the authors compute the bending of light (and massless scalar particles) by the sun, see also \cite{BjDoHo2,BaiHua}. At leading order, the amplitude reads
\begin{equation}
\mathcal{M}^{\text{tree}}\propto (M\omega)^2G/t,
\end{equation}
where $M$ is the mass of the sun and $\omega$ the massless particle energy in the sun reference frame. Applying Eq. (\ref{pot}), one finds
\begin{equation}
V(r)^{\text{tree}}\propto -GM\omega/r,
\end{equation}
reminiscent of the Newtonian potential between two masses. Note that here, the energy denominator in Eq. (\ref{pot}) is $4M\omega$. The 1-loop calculation gives two types of contributions
\begin{equation}
\mathcal{M}_{cl}^{\text{1-loop}}\propto (M\omega)^2G^2\frac{M}{\sqrt{t}} \;\;\; , \;\;\; \mathcal{M}_{qu}^{\text{1-loop}}\propto (M\omega)^2G^2\ln \left(-\frac{t}{E^2}\right),
\end{equation}
where $E=M$ or $E=\mu$, and $\mu$ is an arbitrary mass scale. As explained in \cite{BjDoHo}, the first contribution, whose Fourier transform is proportional to $(\hslash)^0G^2r^{-2}$, is of classical nature and corresponds to one of the post-Newtonian corrections to the leading-order potential. The other term is a quantum correction 
proportional to $\hslash$
\begin{equation}
V(r)_{qu}^{\text{1-loop}}\propto \frac{\hslash M\omega G^2}{r^3}. 
\end{equation}
At large distances, these quantum corrections are negligible compared to the classical contributions. The potential found in Ref. \cite{BjDoHo} gives a bending angle in agreement with measurements, showing the adequacy of low-energy quantum gravity and semi-classical calculations at large enough distances.\\

Finally, note that dimensional analysis allows contributions to the potential such as $(M\omega)^2G^2/\hslash r$. These contributions are clearly problematic for the classical limit, and are absent in \cite{BjDoHo}. In Ref. \cite{DonTor}, the authors showed that in the case of the gravitational interaction between two masses, such contributions can arise from individual Feynman diagrams. However, a cancellation occurs once the diagrams are properly summed.

\section{The graviton-graviton potential}\label{secgrav}
\subsection{At tree level}
The graviton-gravition scattering amplitude has been reported in several articles, see for instance \cite{DeKrSc,Ber,GrNiWu}. Accounting for the convention used in \cite{DeKrSc}, where $i \mathcal{M}$ is written $\mathcal{M}$, the amplitude reads
\begin{equation}
\mathcal{M}^{\text{tree}}_{++;++}=\left( \frac{\kappa}{2}\right)^2 \frac{s^3}{tu},\label{Atree}
\end{equation}
where $s$, $t$ and $u$ are the usual Mandelstam variables, $s=(P_{1}+P_{2})^{2}$, 
$t=(P_{1}-P_{3})^{2}=q^2$, 
and $u=(P_{1}-P_{4})^{2}$.
The $+$ denotes the helicity of the gravitons, and we use the physical helicity\footnote{Another convention is to consider all particles in the amplitude as outgoing. In this case, particles $1$ and $2$ have their helicity reversed.}. At tree level, helicity combinations other than $(++;++)$ are zero. As explained in Sec. \ref{secpot} the long-distance potential is proportional to the Fourier transform of the amplitude in the limit $t\rightarrow 0$. Dividing by $s$ due to the energy factors in Eq. (\ref{pot}) and using the relation
\begin{equation}
\lim_{t\to 0} u=-s,
\end{equation}
for massless particles, we obtain the leading-order potential
\begin{equation}
V^{\text{tree}}(r)=-\left( \frac{\kappa}{2}\right)^2\int \frac{d^3\vec{q}}{(2\pi)^3}e^{i\vec{r}.\vec{q}}\frac{s}{\vec{q}\,^2}=-\left( \frac{\kappa}{2}\right)^2\frac{s}{4\pi r}.\label{potgrav}
\end{equation}
Since $s \geq 0$, the potential is always attractive, as expected.

\subsection{The 1-loop result}
\begin{figure}[!h]
\begin{subfigure}{0.31\textwidth}
\includegraphics[width=8.0cm]{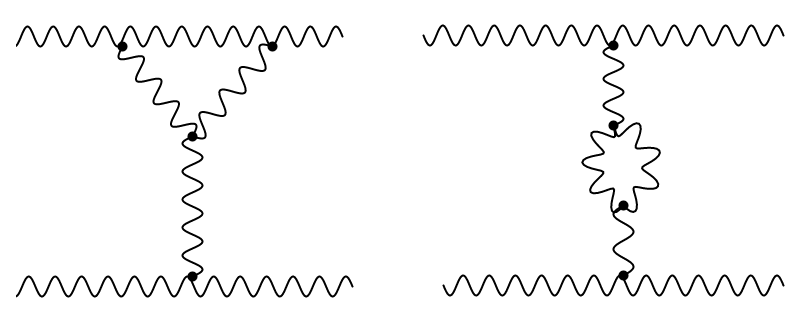}
\caption{}\label{oneLoop}
\end{subfigure}
\hspace*{\fill} 
\begin{subfigure}{0.31\textwidth}
\includegraphics[width=4.2cm]{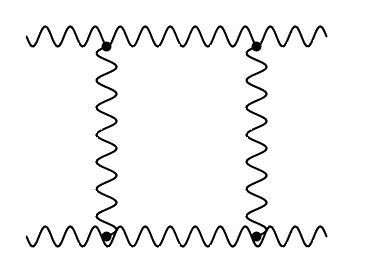}
\caption{}\label{oneLoop2}
\end{subfigure}
\caption{(a): Examples of 1-loop corrections to the one-graviton exchange. The wavy lines represent gravitons. (b): Example of a two-graviton exchange.}
\end{figure}

The potential given by  Eq. (\ref{potgrav}) can receive two kinds of corrections: from higher order terms in Eq. (\ref{leff}) and from loop calculations in Einstein gravity\footnote{In the context of effective field theory, Einstein gravity refers to the Einstein-Hilbert Lagrangian, linear in $R$. Higher order terms in Eq. (\ref{leff}) are not used for the 1-loop calculation.}. We already mentioned that the formers are negligible at sufficiently large distance, and in particular at the graviball event horizon. Power counting indicates that an n-loop amplitude in Einstein gravity is of the same order as $\mathcal{L}_n\propto R^{n+1}$ \cite{Burg,DonTor}. Here and in the following, we use the notation $R^n$ to denote the terms formed by the appropriate contraction of $n$ factors chosen among $R$, $R_{\mu\nu}$ and $R_{\mu\nu\alpha\beta}$. Power counting being sometimes misleading, the corrections generated by the 1-loop amplitude must be checked explicitly. However, we are only interested in a class of 1-loop corrections corresponding to the one-graviton exchange, illustrated in Fig. \ref{oneLoop}. The two-graviton exchange Fig. \ref{oneLoop2}, and more generally the multi-graviton exchange is already included in the formalism of Ref.~\cite{AdBiVe} and effectively included in our semi-classical treatment, see Fig.~\ref{grib} and the corresponding discussion. The main difference between Figs. \ref{grib} and \ref{oneLoop2} is that in the former, the horizontal internal lines are on-shell or nearly on-shell.\\

While the 1-loop amplitude has been long known \cite{DunNor}, the 2-loop calculation has been published only recently \cite{AbCoIt}. In pure gravity, the 1-loop result for $\mathcal{M}_{++;++}^{1-loop}$ is given in Ref. \cite{DunNor} by the involved equations (4.4) and (4.8)\footnote{In Ref. \cite{DunNor} the convention used is that all particles in the amplitude are outgoing. Then, helicities of particles $1$ and $2$ are reversed compared to our convention.}. Note also that while the amplitudes $\mathcal{M}_{+-;++}^{1-loop}$ and $\mathcal{M}_{--;++}^{1-loop}$ are not zero at {1-loop}, in contrast to the tree-level case, they do not contain non-analytic terms and consequently do not contribute to the long range potential.\\

While the 1-loop diagrams with at least one massive external particle contain in general a term in $1/\sqrt{t}$,\footnote{The origin of the $1/\sqrt{t}$ term is the so-called mixed loop, a loop with both massive and massless propagators.} whose Fourier transform gives a contribution $\propto 1/r^2$ to the potential, no such terms are present in the four-graviton case. Ignoring the infrared pole, which cancels at the cross section level against real emission diagrams \cite{DonTor2}, the pure gravity amplitude $\mathcal{M}_{++;++}^{\text{1-loop}}$ contains only the following non-analytic terms
\begin{equation}
2F\left(\frac{\ln(-\tilde u)\ln(-\tilde t)}{\tilde u\tilde t}+\frac{\ln(-\tilde s)\ln(-\tilde t)}{\tilde s \tilde t}\right), \label{1loop1}
\end{equation}
with $F$ given by:
\begin{equation}
F=\frac{i\tilde s\tilde t\tilde u\kappa^2 (4\pi)^{\epsilon}}{4(4\pi)^2}\frac{\Gamma^2(1-\epsilon) \Gamma(1+\epsilon)}{\Gamma(1-2\epsilon)}\mathcal{M}^{\text{tree}}_{++;++},
\end{equation}
where $\mathcal{M}^{\text{tree}}_{++;++}$ is given in Eq. (\ref{Atree}), $\epsilon<0$ and vanishingly small, and $\Gamma$ is the gamma function. The tilde signals the definition of the Mandelstam variables used in \cite{DunNor}, since the authors work in the unphysical regime where $\tilde s<0$. In this case, the relation between Mandelstam variables is 
$\tilde t+\tilde u=\tilde s.$
In the limit $|\tilde t| \ll |\tilde s| \simeq |\tilde u|$, the two terms in Eq. (\ref{1loop1}) are identical and give the contribution to the amplitude:
\begin{equation}
\mathcal{M}_{++;++}^{\text{1-loop}}=-\frac{s^3\kappa^4}{4(4\pi)^2}\ln(s)\frac{\ln\left(\vec{q}\,^2\right)}{\vec{q}\,^2}.\label{1simp}
\end{equation}
Using Eqs. (\ref{pot}) and (\ref{1simp}), the correction to the leading-order potential would be:
\begin{align}
V(r)^{\text{1-loop}}&=-\frac{8s^2\ln(s) G^2}{\pi}\frac{[\gamma_E+\ln (r)]}{r} \nonumber \\
&=\frac{2s\ln(s)G}{\pi}[\gamma_E+\ln (r)]V(r)^{\text{tree}},
\label{v4g_1loop}
\end{align}
where $\gamma_E$ is the Euler constant. Restoring momentarily $\hslash$ and $c$, we find
\begin{equation}
V(r)^{\text{1-loop}}=\frac{2s\ln(s)G}{\hslash c \pi c^4}[\gamma_E+\ln (r)]V(r)^{\text{tree}}.\label{pothbar}
\end{equation}
However, such contribution with a $\hslash$ in the denominator typically originates from the two-graviton exchange \cite{DonTor}, which, accordingly to the previous discussion, should not be taken into account in the semi-classical treatment.\\

The remaining terms in equation (4.8) of Ref. \cite{DunNor} are:
\begin{equation}
\frac{F\ln^2(t/u)(t+2u)(2t+u)(2t^4 + 2t^3u - t^2u^2 + 2tu^3 + 2u^4)}{s^8}
\end{equation}
and
\begin{equation}
-\frac{F\ln(t/u)(t-u)(341 t^4 + 1609 t^3u + 2566 t^2u^2 + 1609 tu^3 + 341 u^4)}{30s^7}.
\end{equation}
They give negligible quantum corrections to the potential
\begin{equation}
V(r)^{\text{1-loop}}\propto G^2\hslash^{n+1}/r^{n+3},
\end{equation}
where $n>0$. Outside the pure gravity sector, the loops can contain other particles like photons. These contributions give also negligible quantum corrections, with at least one power of $\hslash$ in the numerator.\\

All in all, in contrast to the case with at least one massive external particle, the 1-loop contribution to the graviton-graviton potential does not yield any classical corrections. The quantum corrections being small, the leading order potential, Eq. (\ref{potgrav}), is expected to provide accurate results.

\section{The $2\rightarrow 2$ graviton interaction: Simulation and results}\label{secres}
We consider the case of two gravitons with identical energy $\omega$ and initial 4-momenta $P_1=(\omega,-\omega,0,0)$ and  $P_2=(\omega,\omega,0,0)$, where the non-zero component of the 3-momenta is on the $x$ axis. 
Then, Eq. (\ref{potgrav}) reads
\begin{equation}
V(r)=-\frac{8G\omega^2}{r}, \label{pot2grav}
\end{equation}
We first investigate the academic case where the gravitons carry the ultra-Planckian energy corresponding to the mass of Earth, $\omega/c^2=5.9\times10^{24}$ kg. 
This could be useful in the context of more realistic studies, where the ultra-Planckian graviton represents a collimated flux of a large number of gravitons with energy $\omega/N$. A case with a smaller $\omega$ value will be discussed in Sec.~\ref{secdis}.\\

\begin{figure}[!h]
\center
\includegraphics[width=12.0cm]{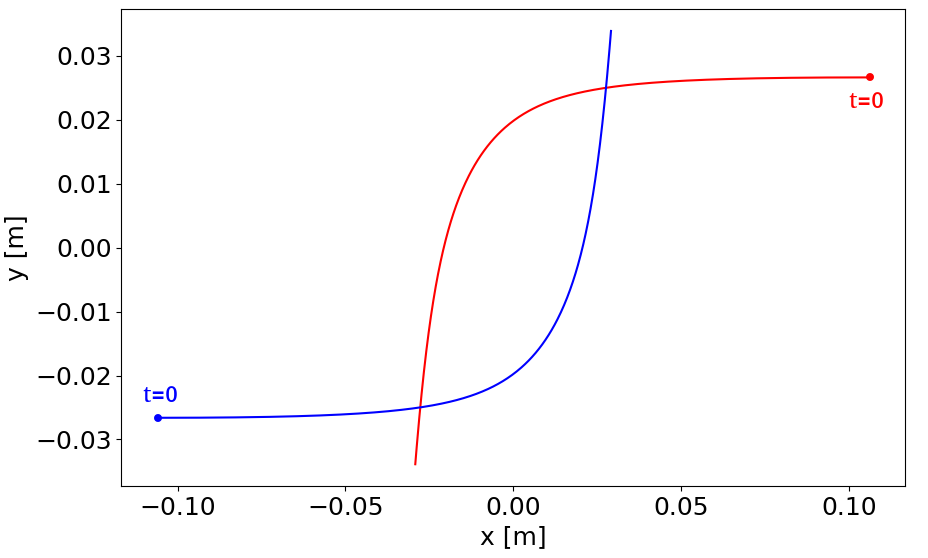}
\includegraphics[width=12.0cm]{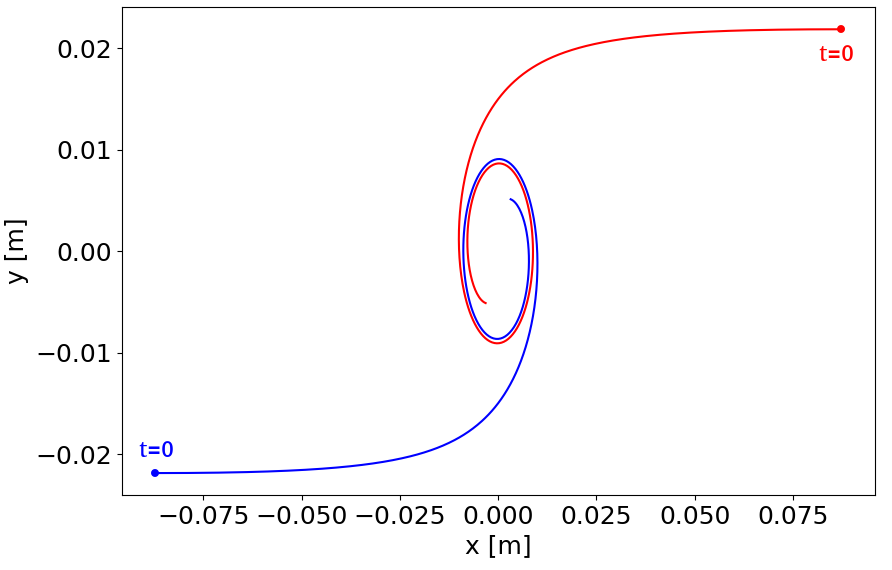}
\vspace{-0.5cm}
\caption{Gravitons trajectories represented by the blue and red lines. In the top panel, the initial impact parameter is larger than the critical value $b_{\text{hor}}$ and the gravitons scatter off each others. In the bottom panel, $b<b_{\text{hor}}$ and the gravitons start to spiral around each others, thereby forming a bound object. Videos associated to these trajectories are available in the arxiv ancillary files.\label{traj}}
\end{figure}

Using the Runge-Kutta method, we solve numerically the relativistic equations of motion, a system of first order (non-linear) differential equations 
\begin{eqnarray}
\frac{dx}{dt}&=&v_x \label{releq1}\\ 
\frac{dy}{dt}&=&v_y \label{releq2}\\ 
\frac{dv_x}{dt}&=&\left( F_x-v_x(v_xF_x+v_yF_y)\right)/\omega \label{releq3}\\
\frac{dv_y}{dt}&=&\left( F_y-v_y(v_xF_x+v_yF_y)\right)/\omega \label{releq4}
\end{eqnarray}
where $\overrightarrow{F}=-\overrightarrow\nabla V(r)$, $F_x=F\cos\varphi$, $F_y=F\sin \varphi$, and $\varphi$ is the angle with the $x$ axis.
We define the impact parameter $b$, to be the initial separation on the $y$ axis, and we choose the initial separation on the $x$ axis to be significantly larger than $b$. At large impact parameter, the interaction causes only a deflection of the gravitons, as illustrated in the top panel of Fig. \ref{traj}. 
Reducing the value of the impact parameter one reaches the critical value $b_{\text{hor}}$ where the gravitons can no longer escape their mutual attraction. The index ``$\text{hor}$" stand for horizon, by analogy with the event horizon of a black hole. As mentioned before, in this region the curvature is still small enough for semi-classical calculations to apply. The trajectory obtained in this case is shown in the bottom panel of Fig. \ref{traj}, where we observe the gravitons spiraling around each other. Videos associated to these trajectories are available in the  arxiv ancillary files. With the chosen input energies, we find that $b_{\text{hor}}\sim 4$ cm.\\

The validity of our calculations stops at small $r$, where the higher orders in Eq. (\ref{leff}) will start to contribute significantly. It is however not possible to determine precisely at which value of $r$ they occur, since it depends on the unknown constants $c_i$. In the case of external particles of mass $m_1$ and $m_2$, the correction to the potential given by the $R^2$ terms is \cite{Ste}:
\begin{equation}
V(r)=-\frac{\kappa^2}{32\pi}\frac{m_1m_2}{r}\left( 1+\frac{1}{3}e^{-M_1r}-\frac{4}{3}e^{-M_2r} \right),\label{potR2}
\end{equation}
where
\begin{align}
M_1^2=\frac{1}{(3c_1+c_2)\kappa^2}; ~~~~~
M_2^2=\frac{2}{c_2\kappa^2}.
\end{align}
In the case of four external gravitons, we expect corrections similar to those in Eq. (\ref{potR2}), since they originate from the graviton propagator. Because the constants $c_1$ and $c_2$ are 
essentially unknown (the limit from laboratory tests of gravity at short scales is  $c_{1,2}<10^{56}$ \cite{DoMiSh}) it is not possible to determine for which value of $r$ these corrections are important. But any reasonable value of the $c_i$ constants will give a negligible contribution; in particular for $c_{1,2}\sim 1$, these corrections start to contribute at $r=r_{\text{rep}}\sim 10^{-35}$m.\\

It is worth noting that we do not expect the conclusion of our study to be changed by the unknown small-distance behavior. Let us consider the situation where higher orders give significant contribution at $r\sim 10^{-35}$m ($c_{1,2}\sim 1$), and where the separation between the gravitons has reached this value. If the small-distance potential is strongly attractive, the two gravitons will still be bound together, and the size of the system will be even smaller than expected by semi-classical calculations. If the small-distance potential is weakly attractive or even repulsive, as exemplified by the third term in the r.h.s of Eq. (\ref{potR2}), the two gravitons will move away, increasing the distance between them. Doing so, the system will be back in a region where the physics is dictated by the potential Eq. (\ref{pot2grav}). This will happen at a distance much smaller than the event horizon, found to be $b_{\text{hor}}\sim 4$ cm  for our academic example. In this case the size of the system will fluctuate between $10^{-35}$ m and $4$ cm.

\section{Toward a many-particle graviball}\label{secreal}
 situations involving more gravitons (and particles), can be classified into four different cases depending on the nature of the two initial colliding objects
\begin{enumerate}
    \item Two collimated beam of gravitons with $r_b \leq L\ll b$, where $r_b$ is the beam radius, $L$ is the length of the graviton bunch and $b$ is the impact parameter.  
    \item Two energetic gravitons producing a large number of particles, similarly to the ideas developed in Refs.~\cite{DvGoIs,AdBiVe} and references therein.
    \item One graviton and one graviball.
    \item Two graviballs.
\end{enumerate}
The first case is the simplest. Note that a collimated beam of gravitons, even for $L\sim r_b$, is not a graviball since gravitons propagating in the same direction do not interact. Indeed, in this case the center of mass energy in Eq.~(\ref{potgrav}) is zero. As long as $r_b\ll b$, the left coming gravitons will react to the bunch of right coming gravitons as if it were one energetic graviton, and the dynamics will be similar to the two-graviton case of Sec.~\ref{secres} (neglecting gravi-strahlung emissions at large angle).\\

The second case is the most interesting but also the most complicated. Here the goal of the simulation would be to investigate the dynamics of the graviball formation, due to the showering process leading to the production of a large number of particles. In Ref.~\cite{AdBiVe}, this process is compared to the parton showers in QCD. Indeed, we believe that, while complicated, it is feasible to implement such processes using Monte-Carlo techniques developed for QCD. The results obtained in \cite{DvGoIs,AdBiVe} are limited by the complexity of the $2\rightarrow N$ amplitude. The phase space integration is not performed and the interactions among the $N$ produced particles are not explicitly addressed\footnote{In Ref.~\cite{DvGoIs} the black hole is described as a Bose-Einstein condensate.}. The advantages of a Monte-Carlo simulation is the possibility to overcome these difficulties and to access to the space-time dynamics of the graviball formation. Of course, numerical simulations are limited by the time-consuming treatment of a large number of particles. In \cite{AdBiVe}, the mean number of particle is estimated to be
\begin{equation}
\langle N\rangle \sim \alpha_G= \frac{Gs}{\hslash c^5}=6.7\frac{s}{1\text{GeV}^2}\times 10^{-39},
\end{equation}
where $\alpha_G$ is the adimentional coupling constant. In this case, the particles have an energy corresponding to the Hawking temperature
\begin{equation}
    E_H=\frac{\sqrt{s}}{\langle N\rangle}\sim \frac{\sqrt{s}}{\alpha_G}.
\end{equation}
For our academic case, where each graviton carries an energy equivalent to the mass of Earth, $\langle N\rangle \sim 10^{64}$ and $E_H\sim 10^{-13}$ GeV. Clearly, such large $\langle N \rangle$ system cannot be numerically simulated. However, in the next section we will discuss the case of initial gravitons with $\omega \sim 10^{19}$ GeV (the corresponding graviball mass is $M\sim 10^{-6}$ kg). For this kinematics, $\langle N\rangle\sim 100$; a reasonable number for a Monte-Carlo simulation. It is this particular case that we will address in a future work. The stability of this object will be discussed in the next section, where we consider the possible role played by graviballs in the resolution of the missing mass problem. In fact, graviballs may be unstable but long lived.\\

The third case is interesting because knowing the probability for a graviton to be trapped by the graviball is directly relevant to the question of the graviball stability. If the graviball is at equilibrium (at the critical value $\frac{\alpha_G}{\langle N\rangle}\sim1$ \cite{DvGoIs,AdBiVe}), the approximation of a fixed particle number could be made. Then, a simplified simulation would consist in considering only elastic interactions among $N=\langle N\rangle$ particles.\\

The fourth case can be viewed as a generalization of the third one. Here, the physics should be similar to the collision of two black holes. This case will not contribute to the understanding of the graviball formation and stability, but could play a role for the phenomenology.

\section{Discussion}\label{secdis}
\begin{figure}[!h]
\center
\includegraphics[width=10.0cm]{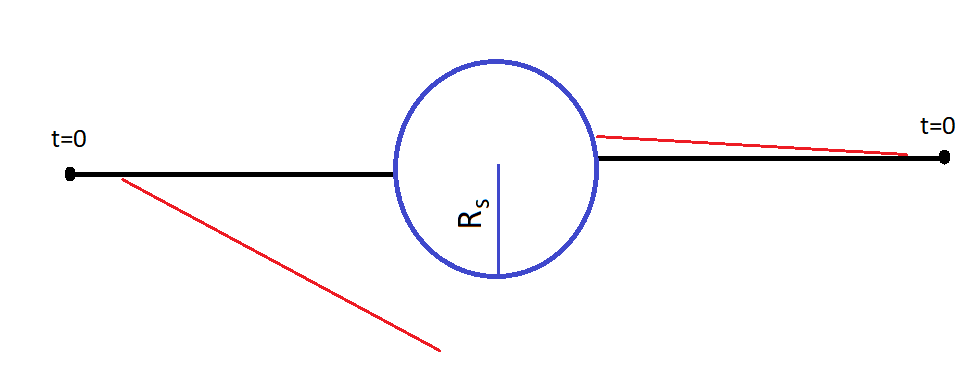}
\vspace{-0.5cm}
\caption{Scattering of two energetic gravitons, resulting in the formation of a graviball of radius $R_s$. The red lines represent soft emissions. The emission in the left will contribute to the energy loss, leading to the reduction of the Schwarzschild radius $R_0\rightarrow R_s<R_0$, where $R_0$ would have been the Schwarzschild radius in the absence of emission.\label{emi}}
\end{figure}

Because we investigated the simplest case of two gravitons of equal 4-momenta, certain aspects of the graviball remain to be investigated, requiring in particular a study of the second and third cases described in the previous section. Before discussing the stability of the graviball, we explain why we do not expect the conclusion obtained in Sec.~\ref{secres} to be significantly changed in more complex situations. At impact parameter $b\gtrsim b_{\text{hor}}\sim R_s$, we expect mainly elastic $2 \rightarrow 2$ scatterings, accompanied by some gravi-strahlung emissions. These emissions reduce the graviton energy and the intensity of its interaction. Consequently, we expect the deflection angle to be slightly reduced. If $b$ is equal or slightly smaller than the limit value $b_{\text{hor}}$, we should observe trajectories with a large deflection angle, instead of the formation of a graviball, again because of the energy loss. For $b$ sufficiently smaller than $b_{\text{hor}}$, we expect inelastic $2\rightarrow N$ scatterings, leading to the graviball formation. Note that the gravi-strahlung emissions prior to the inelastic scattering will not necessarily contribute to the energy loss, as illustrated in Fig.~\ref{emi}. It is in particular the case for collinear emissions.\\

For the sake of coherence, it is interesting to ask why, or under which condition, the (gravi-strahlung) emitted graviton can escape from the energetic graviton potential well. In order to answer this question, we write the potential Eq.~(\ref{potgrav}) for the gravitons 4-momenta $P_L=\omega_L(1,1,0,0)$ and $P_e=\omega_e(1,\cos\theta,\sin\theta,0)$:
\begin{equation}
    V(r)=-\frac{4G\omega_L\omega_e(1-\cos\theta)}{r}.\label{potemi}
\end{equation}
Here, the subscripts $L$ and $e$ stand for leading graviton and emitted graviton, respectively. It is clear that the intensity of the interaction, or equivalently the Schwarzschild radius seen by the emitted graviton, goes to zero when $\omega_e\rightarrow 0$. Then, if the initial distance between the leading and emitted gravitons is larger than $R_s$, the latter will escape (it is actually emitted). It is also interesting to note that, if the gravitational interaction is repulsive at small distances $r<r_{\text{rep}}$ (see the discussion in Sec.~\ref{secres} associated to Eq.(\ref{potR2})), an event horizon exists only if the Schwarzschild radius associated to the attractive potential (\ref{potemi}) is larger than $r_{\text{rep}}$. If $R_s < r_{rep}$, there is no event horizon and the emitted graviton will be able to escape even if the initial distance between the two particles is zero. To summarize, only soft enough gravitons can be emitted. This result, in apparent contradiction with the fact that a particle inside a black hole cannot escape, independently of its energy, is a specificity of microscopic systems (made of few particles). Indeed, in the gravi-strahlung case
\begin{equation}
    \lim_{\omega_e\to 0} s=\lim_{\omega_e\to 0} (P_L+P_e)^2=P_L^2=0,
\end{equation}
and $R_s$ can reach the value zero. This is not true anymore for $N$ particles (taken to be at the same space-time coordinate), since
\begin{equation}
    \lim_{\omega_e\to 0} (P_1+...+P_N+P_e)^2=(P_1+...+P_N)^2\neq 0,
\end{equation}
except if the $N$ particles propagate in the same direction. With the growing number of particles, the contribution of $P_e$ to $s$ becomes more and more negligible, and the Schwarzschild radius seen by the emitted graviton becomes independent of $\omega_e$, in agreement with classical expectations.\\

Finally, and before discussing the graviball as a potential solution to the missing mass problem, we discuss its stability. We do not expect decays similar to what happens to QCD's bound states, for several reasons. One is that massive enough graviballs are analogous to the black holes discussed for instance in \cite{DvGoIs,AdBiVe}. Clearly, those are stable. We do not expect special decay mechanism to emerge for lighter graviballs. Another reason is that there are several differences between QCD and gravity: Not all particles have a color charge, while all particles carry a 4-momentum; the color charge can be screened, not the 4-momentum; and gravity is always attractive (ignoring the behavior below the Planck scale). Consider the decay $D^+\rightarrow \mu^-\mu^+\pi^+$. It can be written in term of the quark constituents $c\bar{d}\rightarrow \mu^-\mu^+ u\bar{d}$, where the two escaping muons have no color charge. In the decay $B^-\rightarrow K^-D^0$, or correspondingly $b\bar{u}\rightarrow \bar{u}c+s\bar{u}$, the colored quarks are organized in color singlet meson, escaping the QCD long range potential $V(r)\propto r$. Due to the differences just discussed between QCD and gravity, these situations are not possible in gravity, and the classical expectation that the energy localized within its Schwarzschild radius will stay there is a good approximation.\\

Despite the previous arguments supporting the graviball stability, a semi-classical expectation is that black holes/graviballs should radiate energy (evaporation), and thus be long lived rather than stable (unless the amount of energy captured by the black hole/graviball is larger than the amount of radiated energy). Several evaporation mechanisms are possible beside Hawking's radiation. In particular, a small number of gravitons could be organized in a configuration such that the resulting force $\vec{F}$ on a test graviton would be aligned with its 3-momentum, leaving its acceleration unchanged. The probability to reach such a configuration should decrease with the number of particles, implying a faster evaporation for lighter gaviballs. This effect is straightforward to study numerically and will be included in our next article.\\

Our results for the two-graviton case readily suggests two candidates for the missing mass problem:
\begin{enumerate}
\item Heavy graviballs.
\item Light graviballs.
\end{enumerate} 
For the first candidate, the situation is similar to that of black holes. Primordial black holes with masses in the range of asteroid masses 
are still candidates to explain the missing mass problem, while lighter or heavier black holes have been ruled out~\cite{Montero-Camacho:2019jte}.
Each black hole, or graviball, generates a large gravitational interaction, and the amount of missing mass requires only a small quantity of these objects. As an illustration, we compute the effect of the two-graviton graviball studied in Sec. \ref{secres}, on the light of a distant galaxy. Here, the photon interacts coherently with the whole 2-graviton system of total 4-momentum $P=(\sqrt{s},0,0,0)$, i.e. of genuine mass $M_g=\sqrt{s}$. One of the main results of Ref. \cite{BjDoHo} being that, except for negligible quantum corrections, the gravitational interaction is spin independent, we can use the expression obtained by the authors for the bending of light by a massive object:
\begin{equation}
\theta\simeq \frac{8GM_g}{dc^2}+\frac{15\pi G^2M_g^2}{d^2c^4},
\end{equation}
where the constant $c$ has been restored for clarity. For distance of closest approach of the light ray e.g. $d=10^3$ m, the first term of this equation gives an angle of $7.2$ arcsec, which is approximately 4 times the value obtained for the sun (measured at the sun surface $d\simeq 6\times 10^8$~m). The second term gives a negligible contribution of $1.8\times10^{-4}$ arcsec. Two distinct scales appear; the size of the graviball, here a few centimeters, and the maximal distance to the graviball where the effects on light are measurable, around $10^4$~m.\\

In case 2, the graviball has been formed by two gravitons with Planckian energies.\footnote{Our formalism may not apply at smaller scale, except if an hypothesis on the small distance behavior of gravity is made.} The event horizon $b_{\text{hor}}$ can be estimated by a formula similar to the Schwarzchild radius:
\begin{equation}
b_{\text{hor}} \simeq \frac{8GM_g}{c^2}.\label{horval}
\end{equation}
Eq. (\ref{horval}) gives the correct order of magnitude as we can check with $M_g\simeq 6\times 10^{24}$ kg yielding $b_{\text{hor}} \simeq 3$ cm, in agreement with the value found in Sec. \ref{secres}. As an example of light graviball, we take $M_g\sim 10^{-6}$ kg, or correspondingly $\sqrt{s}\sim 10^{19}$ GeV. In this case $b_{\text{hor}}\sim 10^{-32}$ m, and the semi-classical calculations are still valid for reasonable values of the $c_i$ constants, see the discussion at the end of Sec. \ref{secres}. This example shows that graviball can be light enough to  escape direct detection, and a large amount of light graviballs, about $10^{50}$, could explain the missing mass problem. For a uniform distribution of light graviballs in the galaxy, this number corresponds approximately to a density of $10^{-15}$~m$^{-3}$. The size of a light graviball
being necessarily small, it could be interpreted, e.g., as a light particle. In fact, Wheeler had speculated on the possible relation between geons and elementary particles \cite{Whe}.

\section{Conclusion}\label{conclusion} 
We have used the formalism of low-energy quantum gravity to study the possible existence of graviballs in the simplifed case of two graviton constituents. Our simulation relies on the potential extracted from the 4-graviton amplitude, and on the numerical resolution of the relativistic equations of motion, thereby providing the space-time dynamics of the graviball formation. While cases including more than two gravitons need to be studied, the conclusion obtained in this article, namely the possibility for gravitons to form a bound system, should not change. Studies of black hole formation at ultra-Planckian energies, e.g. \cite{DvGoIs,AdBiVe}, support this expectation. This opens the prospect of graviballs being a solution to the galactic and cluster missing mass problem. In that context, we computed the gravitational lensing caused by graviballs in order to assess their detectability.

\section*{Acknowledgments}
We would like to thank J. Donoghue and K. Krasnov for additional explanations on \cite{DonTor} and \cite{DeKrSc}, respectively. We also thank O. Castillo Felisola for bringing to our knowledge Ref. \cite{BjDoHo}.  We thank I. Schmidt and B. Terzić for their comments on the manuscript. B.G. acknowledges support from Chilean FONDECYT Iniciaci\'on grant 11181126. B.G. is supported by ANID PIA/APOYO AFB180002 (Chile).

\begin{appendices}

\section{Geons and graviballs\label{apgeon}}
The term geon (Gravitational Electromagnetic entity) was coined by Wheeler when he studied a system of electromagnetic waves held together by their gravitational attraction \cite{Whe}. The case of gravitational waves, called gravitational geon, was studied first by Brill and Hartle \cite{BriHar}. Gravitational geons have been revisited and discussed in several papers, e.g. \cite{AndBri,PerCoo}. While here, we study graviballs using a quantum formalism, the gravitational geon is a classical object described by 
general relativity. Calculations are non-trivial, and the stability of the gravitational geon  is questioned in Ref. \cite{PerCoo} due to an inconsistency found between the final result and the initial hypothesis on time-scale evolution. 

Our (quantum) calculations for the graviball offer
several advantages and improvements compared to classical calculations. First, we do not need to solve the Einstein equations. Furthermore, the effective field theory formalism allows a better control on the validity of the performed calculations, see for instance the discussion on the small-distance behavior at the end of Sec.~\ref{secres}. This could help in the debate on the curvature singularity for gravitational geons. In some articles, e.g. \cite{AndBri}, it is said that the curvature has no singularity. However, in Ref. \cite{PerCoo} the authors claim that space-time cannot be taken as singularity-free. Our calculations  agree with the latter statement. Another advantage is that we do not employ the time averaging used for gravitational geons. In our calculation, the potential energy is computed at each step of the evolution, allowing a realistic dynamical study of the graviball.

\end{appendices}


\end{document}